\documentclass[usenatbib]{mn2e}
\usepackage{graphicx, txfonts, natbib}
\usepackage{epsf}
\usepackage{amssymb}
\usepackage{epstopdf}
\usepackage{longtable}
\newcommand{\msun}{\mbox{M$_{\odot}$}}

\DeclareMathAlphabet{\mathsc}{OT1}{cmr}{m}{sc}
\def\testbx{bx}%
\DeclareRobustCommand{\ion}[2]{%
\relax\ifmmode
\ifx\testbx\f@series
{\mathbf{#1\,\mathsc{#2}}}\else
{\mathrm{#1\,\mathsc{#2}}}\fi
\else\textup{#1\,{\mdseries\textsc{#2}}}%
\fi}

\newcommand{\Feii} {\ion{Fe}{ii}}

\begin{document}

\title[Type II-P supernovae as standardised candles: improvements using near infrared data]
  {Type II-P supernovae as standardised candles: improvements using near infrared data\thanks{Based on data obtained with telescopes from 
Las Campanas Observatory and the 
University of Arizona.}}

\author[K. Maguire et al.]
	  {K.~Maguire,$^{1} \thanks{E-mail: kmaguire11@qub.ac.uk}$
   R.~Kotak,$^1$ S. J.~Smartt,$^1$ A.~Pastorello,$^1$ M.~Hamuy,$^2$ F.~Bufano$^3$ \\
      $^1$Astrophysics Research Centre, School of Maths and Physics, Queen's University Belfast, Belfast BT7 1NN, UK\\
        $^2$Universidad de Chile, Departamento de Astronom\'ia, Casilla 36-D, Santiago, Chile\\
      $^3$INAF, Osservatorio Astronomico di Padova, Vicolo dell'Osservatorio 5, I-35122, Padova, Italy\\
      }

\maketitle

\begin{abstract}
We present the first near infrared Hubble diagram for type II-P supernovae to further explore their value as distance indicators. We use a
modified version of the standardised candle method which relies on the tight correlation between
 the absolute magnitudes of type II-P supernovae and their expansion velocities during
the plateau phase. Although our sample contains only 12 II-P supernovae and they are 
necessarily local ($z$ $<$ 0.02), we demonstrate using near infrared $JHK$ photometry that it may be possible
to reduce the scatter in the Hubble diagram to 0.1--0.15 magnitudes.
While this is potentially similar to the dispersion seen for type Ia supernovae, we caution that this
needs to be confirmed with a larger sample of II-P supernovae in the Hubble flow.
\end{abstract}

\begin{keywords}
distance scale -- supernovae: general -- galaxies: general -- cosmology: distance scale
\end{keywords}

\section{Introduction}

Cosmology using type Ia supernovae (SNe) has been established for over a decade since 
the discovery of cosmic acceleration \citep{rie98, per99}. Since these early results, 
distance measurements to type Ia SNe have been estimated to an accuracy of 7 per cent \citep{ast06}. 
Although there is some consensus on the progenitors of type Ia SNe, and the physics
governing these thermonuclear explosions, direct observational constraints are still
lacking. Their use as standardised candles is limited primarily by systematic uncertainties,
but significant effort is being expended in understanding and quantifying the source of
these \citep[e.g.][]{ast06}. Concurrently, several studies have attempted to determine
whether there is any evolution in properties between the low and high-redshift samples
\citep[e.g.][]{riess:99,hook:05} and considered possible correlations with host galaxy 
properties \citep[e.g.][]{gallagher:05,2006ApJ...648..868S}. Complementary methods of 
distance determination out to redshifts roughly comparable to those sampled by type Ia SNe,
would provide a useful and independent check. 

In contrast to type Ia SNe, the progenitors of several types of core-collapse SNe have 
have been unambiguously detected in the local Universe \citep[see][for a review]{2009ARA&A..47...63S}.
 Several point sources have been detected
in pre-explosion images at the location of type II-P SNe with colours 
consistent with those of single red supergiants \citep{sma04, li05,2005MNRAS.364L..33M,2008ApJ...688L..91M, eli09}.
Type II-P SNe are 
characterised by the presence of broad P-Cygni lines of hydrogen in their spectra, while 
an extended plateau of $\sim$ 80--120 days in the light curve gives the class its name. The light curves and spectra of type
II-P SNe have been successfully modelled by several authors \citep[e.g.][]{des08,kas09}.
Although core-collapse SNe exhibit considerable diversity compared with type Ia
SNe, type II-P SNe are arguably the most homogeneous subset of core-collapse SNe at optical \citep{ham03a} and also ultra-violet \citep{gal08,buf09} wavelengths,
and despite their intrinsically lower luminosities, constitute a promising class of
objects that can be exploited to determine
distances. The fact that the progenitor stars are constrained to be red supergiants
between 8--17 \msun\ \citep{sma09} enhances our confidence that the physics
of these explosions are based on firm ground.

Different methods have been investigated to estimate the distance to type II-P SNe, such 
as the original expanding photosphere method \citep{kir74,sch94,ham01} and the more 
recent synthetic spectral-fitting expanding atmosphere method \citep[e.g.,][]{bar04,des08}. 
Both methods rely on high signal-to-noise photometry and spectroscopy, as well as the 
computation of detailed synthetic spectra for each SN as a function of time. 
A much simpler method for using type II-P SNe as standardised candles was introduced by 
\cite{ham02} and \cite{ham03}. This requires much less input data and is based on the strong correlation 
between the expansion velocity and the luminosity of a type II-P SN during the plateau 
phase. 

\cite{nug06} presented refinements to the original method of \cite{ham02} with the aim 
of increasing the ease with which distances to type II-P SNe could be measured at 
cosmological redshifts. 
The most signifcant of these was to determine an extinction correction from the 
\textit{V}-\textit{I} colours at +50 days, rather than at the end of the plateau
phase \citep{ham03}.
They also investigated the effect of using different lines to determine the expansion velocity, and presented an empirical relation -- based on data
available at the time -- in the form of a power law that allows the inference of the 
Fe~{\sc ii} $\lambda$5169 velocity at the epoch of interest, chosen to be +50 days. Applying their method to a sample 
consisting of both local as well as 5 intermediate redshift ($z \le 0.3$) type II-P SNe, 
they found a scatter of 0.26 mag in the \textit{I} band Hubble diagram, similar to 
the dispersion obtained by \cite{ham03} for SNe in the Hubble flow. 

Recently, \citet{poz09} analysed a sample of 40 new and previously published type II-P SNe using 
the standardised candle method. They found a scatter of 18 per cent in distance,
but noticed that 3 SNe deviated significantly from their best-fitting solution, and that 
some of these SNe had faster decline rates. They introduced a criterion based on a weak correlation between the $I$ band decline rate and the deviation from best fit, which rejected 6 SNe including the 3 outliers, which left a final sample of 34 objects, with a scatter of 10 per cent. 
\cite{oli09} also applied the SCM to a sample of 37 type II-P SNe, using a reference epoch of -30 days from the end of the plateau. The scatter of their data in the $B$, $V$ and $I$ was comparable to that of previous studies but with an unexpected small increase in scatter at longer wavelengths. 
Furthermore, \citet{dan09} complied a sample of 15 type II-P SNe from the Sloan 
Digital Sky Survey (SDSS) in the redshift range $0.03 < z < 0.14$. Applying the standardised 
candle method to a combined sample of the SDSS observations and data from the literature, they also found a dispersion comparable to that previously found.
However neither \cite{dan09} nor \cite{oli09} found  fit parameters consistent with those of \cite{poz09}. \cite{dan09} concluded for their sample
that this discrepancy was most likely due to the SDSS SNe being intrinsically brighter, but not showing 
the corresponding increase in ejecta velocities. The early epochs at which the spectra are taken 
for the SDSS sample and the subsequent extrapolation using equation 2 of \cite{nug06} is 
thought to be the main source of this difference in the fit parameters, and highlights the need for reliable measurements 
of the expansion velocities close to the chosen reference epoch (+50 days). \cite{dan09} also find that the {\it ad hoc\/} decline rate culling criterion of \cite{poz09} does 
not remove the outliers from their sample.

\begin{table*}
 \caption{Redshifts, magnitudes and expansion velocities of a selection of type II-P SNe in order of increasing flow velocity.}
 \label{nir_candle}
 \begin{tabular}{@{}lccccccccccccccccccccccccccccccc}
  \hline
  \hline
  SN & 	Heliocentric $cz$&Flow $cz$&\textit{V} &\textit{I}&\textit{J}  &\textit{H} &\textit{K} &  Velocity$_{\Feii}$ &Ref. \\
  	&	 (km s$^{-1}$) &  (km s$^{-1}$)&(mag)  & (mag)   &  (mag)   & (mag)  & (mag)  &  (km s$^{-1}$) &\\
    \hline

SN 2004dj & 131 & 252 $\pm$ 24$^{1}$ & 12.04 $\pm$ 0.03 & 11.40 $\pm$ 0.03 & 10.73 $\pm$ 0.50 & 10.24 $\pm$ 0.30 & ---& 3132 $\pm$ 300 &1, 2\\
SN 2002hh&48&447 $\pm$ 100$^{2}$	&16.35 $\pm$ 0.05&13.68 $\pm$ 0.05&12.30 $\pm$ 0.03&---&11.07 $\pm$ 0.06&4372 $\pm$ 300&3, 4, 5\\
SN 2004et&48&447 $\pm$ 100$^{2}$	&12.90 $\pm$ 0.02&11.95 $\pm$ 0.02&11.46 $\pm$ 0.05&11.10 $\pm$ 0.05&10.88 $\pm$ 0.05&4101 $\pm$ 200&6\\
SN 2005cs&463&	548 $\pm$ 390&14.76 $\pm$ 0.03&13.95 $\pm$ 0.04&13.61 $\pm$ 0.04&13.45 $\pm$ 0.03&13.37 $\pm$ 0.05&1890 $\pm$ 200&7  \\
SN 1999br &960&836 $\pm$ 390	&17.58 $\pm$ 0.05&16.71 $\pm$ 0.05&16.22 $\pm$ 0.10&16.01 $\pm$ 0.20&15.86 $\pm$ 0.50&1545 $\pm$ 300&8, 9\\
SN 1999em& 717&917 $\pm$ 70$^{3}$&13.81 $\pm$ 0.04&13.23 $\pm$ 0.03&12.95 $\pm$ 0.04&12.79 $\pm$ 0.05&12.57 $\pm$ 0.08&3554 $\pm$ 300&5, 10, 11  \\
SN 2005ay& 809 &1086 $\pm$ 390&15.30 $\pm$ 0.06&14.67 $\pm$ 0.02&14.28 $\pm$ 0.10&14.09 $\pm$ 0.30&---&3397 $\pm$ 200&5, 9, 12\\
SN 2003hn&1168 &1389 $\pm$ 390&14.90 $\pm$ 0.05&14.21 $\pm$ 0.04&13.73 $\pm$ 0.05&13.55 $\pm$ 0.03&13.30 $\pm$ 0.05&4263 $\pm$ 200&11   \\
SN 1990E &1241& 1425 $\pm$ 390&15.90 $\pm$ 0.20&14.56 $\pm$ 0.20&13.44 $\pm$ 0.10&13.30 $\pm$ 0.10&---&5324 $\pm$ 300 &8, 13\\
SN 2007aa&1465&2018 $\pm$ 390	&15.86 $\pm$ 0.40&---&14.99 $\pm$ 0.05&14.65 $\pm$ 0.07&---&3011 $\pm$ 260&9\\
SN 2002gd&2674 &	2734 $\pm$ 390&17.52 $\pm$ 0.05&16.78 $\pm$ 0.03&16.70 $\pm$ 0.20&16.30 $\pm$ 0.30&16.20 $\pm$ 0.30&2050 $\pm$ 300&5, 14  \\
SN 1999cr &6055&6539 $\pm$ 390	&18.33 $\pm$ 0.05&17.63 $\pm$ 0.05 &17.40 $\pm$ 0.40&16.85 $\pm$ 0.40&16.62 $\pm$ 0.40&4389 $\pm$ 300&8, 9\\
 \hline
 \end{tabular}
     $^{1}$From the Cepheid distance of \cite{fre01}. 
   $^{2}$From the average distance of \cite{bot09}. 
   $^{3}$From the Cepheid distance of \cite{leo03}.
    REFERENCES -- (1) \cite{vin06} (2) Di Carlo (in prep.); (3) \cite{mei02}; (4) \cite{poz06}; (5) \cite{poz09}; (6) \cite{mag09}; (7) \cite{pas09}; (8) \cite{ham03}; (9) this paper; (10) \cite{leo03}; (11) \cite{kri09}
    (12) Bufano et al. (in prep.) (13) \cite{sch93}; (14) \cite{pas03}
\end{table*}

Given that all previous studies incorporating local and intermediate redshift
samples of type II-P SNe find a scatter of $\gtrsim$ 0.2 magnitudes in the $I$ band,
we investigate here whether further improvements are possible using photometry in the
near-infrared region (NIR). Galactic type dust causes a factor of greater than 5 times 
less extinction in the \textit{H} band than in the \textit{V} band, so one might reasonably
expect a reduced scatter in the Hubble diagram of type II-P SNe. Indeed, it has long been 
known that type Ia SNe are excellent standard candles in the NIR \citep[e.g.][]{meikle:00,kri04,woo08}.
As a first step, we explore the potential benefits of using NIR imaging for local
objects, for which data is currently available. 

\section{Supernova sample}
\label{lum_vel}

\cite{ham01} detailed how the bolometric luminosity of a type II-P SN can be determined from 
its \textit{BVI} photometry, an empirical bolometric correction, total extinction and distance 
to the host galaxy \citep[see also][]{ber09}. He noted that SNe with brighter plateaus have higher expansion velocities. 
Instead of determining the total bolometric luminosity, we used NIR photometry to derive 
a relation between expansion velocity and NIR luminosity. This is done for two 
main reasons: 
(i) the substantially lower extinction in the NIR region compared to optical wavelengths 
implies that the corresponding error on the extinction will have a smaller effect on the fit,
(ii) NIR plateau phase spectra of type II-P SNe contain far fewer lines compared to optical 
regions \citep[e.g.][]{mag09}; thus the NIR magnitudes are presumably 
affected less by variations in line strengths and widths from one SN to another.

We searched the literature for NIR photometry and optical spectroscopy to supplement our own 
data. The optical (\textit{VI} band) magnitudes for each SN were interpolated to +50 days 
using a linear fit. For SN 2007aa, which did not have good coverage at the mid-plateau phase, 
we used the available data at an earlier epoch (+24 days) and scaled this to the magnitude of each of the well observed II-P SNe: 1999em, 2004et and 2005cs at the same epoch. 
The offset at +50 days was obtained for each SNe and the mean value was taken as the magnitude of SN 2007aa. 
In the NIR, the magnitudes that were obtained from the literature were interpolated 
using a linear fit if data points were available within five days of this epoch, or using a quadratic 
interpolation if this was not the case.
The fitting of quadratic interpolation was necessary to account for the pronounced non-linear 
behaviour of the plateau in the NIR bands during the photospheric phase \citep[see][]{mag09}.
For some SNe (SN 1999br, SN 1999cr, SN 2005ay and SN 2007aa), previously unpublished NIR data were analysed in the standard manner to obtain \textit{JHK} magnitudes, which could be interpolated to +50 days based on the criteria detailed above. For SN 1999br, SN 1999cr and SN 2002hh, the $K_{short}$ filter was used, which has a long wavelength cut off at 2.3 $\mu$m. Transformation equations can be applied that convert from the $K_{short}$ to the $K$ band but the coefficients of the conversion are smaller than the errors on the photometry and so have not been applied.

We estimated the expansion velocity by measuring the position of the minimum of the 
Fe~{\sc ii} $\lambda$5169 feature in spectra taken at +50 days for each SN, except for SNe 
1990E, 1999br, and 1999cr, where the expansion velocities were taken directly from \cite{ham03}. They also used the minima of the Fe~{\sc ii} $\lambda$5169 lines and a power-law as described in \cite{ham01} to determine the expansion velocities at +50 days. The spectra used to reanalyse the expansion velocities were obtained from the references 
listed in Table \ref{nir_candle}. When a spectrum at +50 days was not available, we extrapolated the velocities using equation 2 of \cite{nug06}, which is reliable from +9--75 days and adds an uncertainty of $<$ 175 km s$^{-1}$ to the expansion velocity.

The recessional velocities of nearby SNe are affected by the peculiar motion of their host galaxies. 
Ideally, SNe in the Hubble flow ($cz$ $>$ 3000 km s$^{-1}$) would be preferred because the peculiar 
motion of the host galaxies would be small compared to their cosmological redshifts. However, given the sample in hand, we proceeded to correct for peculiar motion as follows: we used the parametric flow model of \cite{ton00}, which has five 
velocity components including a Hubble flow, a constant dipole and quadrupole, and components to 
account for the infall to the Virgo and Great Attractors. The heliocentric velocities of the 
host galaxies were obtained from NED
 and a Hubble constant, $H_0$ of 78.4 km s$^{-1}$ Mpc$^{-1}$ 
is used throughout this paper to ensure consistency with \cite{ton00}. However, the value of $H_{0}$ is only a scaling 
factor and affects neither the slope nor the fit.

We investigated the uncertainty in the recessional
velocities calculated from the flow model of Tonry et al.
(2000) using a sample of galaxies with Cepheid distances from
Freedman et al. (2001). We compared the velocities obtained
from the flow model to those obtained using the Cepheid measurements,
and found the standard deviation between the two
methods to be 342 km s$^{-1}$. This value was added in quadrature
to an uncertainty of 187 km s$^{-1}$, which corresponds to the cosmic
thermal velocity \citep{ton00}, giving a total uncertainty 
for the recessional velocities of 390 km s$^{-1}$.

SN 2004et and SN 2002hh occurred in the nearby galaxy, NGC 6946, which has a heliocentric velocity of 48 km s$^{-1}$ implying that the recessional 
velocity is dominated by peculiar motion and \citet{ton00} noted that their model is uncertain
within $\sim$10 Mpc and particularly for galaxies within $\sim$ 6 Mpc. The flow velocity for these SNe is instead calculated from 
 the average distance to their host complied by \cite{bot09} of 5.7 $\pm$ 0.2 Mpc with $H_0$ = 78.4 km s$^{-1}$ Mpc$^{-1}$. 
For SNe 1999em and 2004dj, a Cepheid distance to the host galaxy 
was available and was used preferentially to derive the redshift. 
The velocities derived from the Cepheid distances are consistent with 
those derived using the flow model to within $<$ 80 km s$^{-1}$. The velocity was calculated for SN 1999em using a distance of 11.7 $\pm$ 1 Mpc \citep{leo03} and for SN 2004dj using a distance of 3.22 $\pm$ 0.15 Mpc \citep{fre01}.

In the original paper of \cite{ham02}, the optical extinction values toward the SNe were taken from the 
literature, while \cite{ham03} determined the optical extinction in the host galaxy using the 
\textit{B-V} or \textit{V-I} colour of the SN at the end of the plateau phase and a corresponding colour 
offset relative to SN 1999em. \cite{nug06} found that using the rest-frame \textit{V-I} colour during 
the plateau phase to estimate the extinction, they could recreate the scatter of \cite{ham03} and 
remove the need for data near the end of the plateau phase. \cite{poz09} allowed for a variable
value of the selective extinction, given that the galaxies that host the SNe studied need not
follow the same extinction law as the Galaxy. They found a best-fitting value of $R_V$ of 1.5 $\pm$ 0.5 
for their sample of 40 SNe. Using the extinction curve of \cite{car89} the value of $R$ in the NIR bands can be obtained. We detail our choice of $R_{V}$ in Section \ref{results}. The optical and NIR magnitudes at +50 days in the Vega system without correction for extinction
are listed in Table \ref{nir_candle} along with 
the heliocentric velocity of the host galaxies, the corrected flow model velocity of each 
SN, and the photospheric expansion velocity of the ejecta at +50 days. 

\begin{figure*}
\includegraphics[width=14cm]{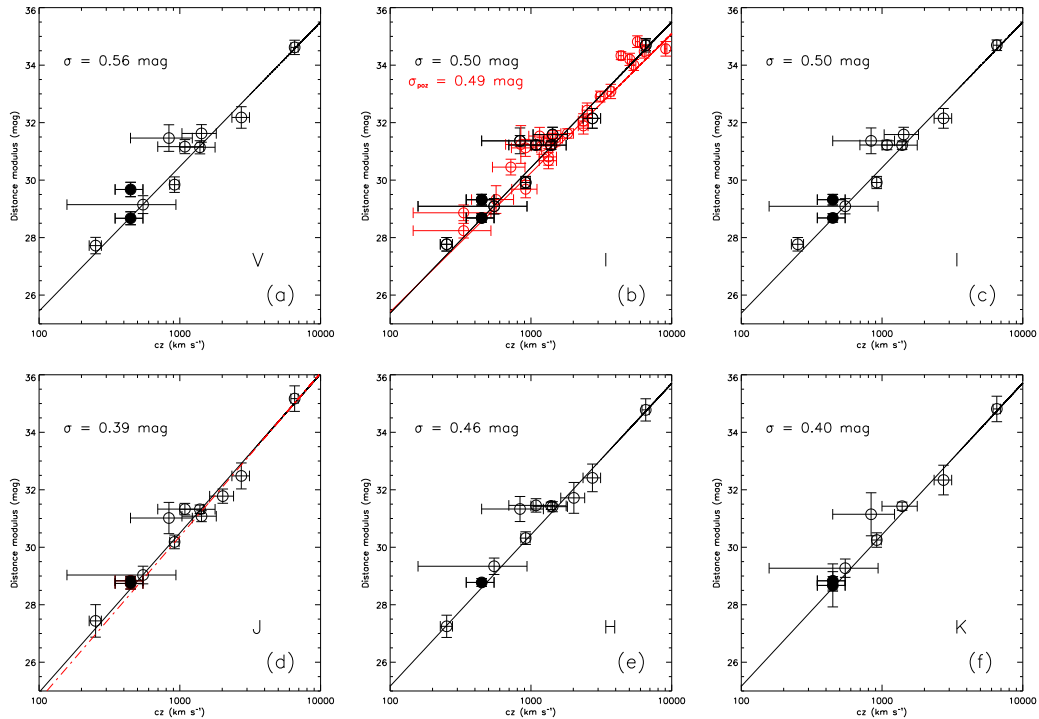}
\caption{Hubble diagrams for the \textit{V}, \textit{I}, \textit{J}, \textit{H} and \textit{K} bands in the panels labelled with the band used. The distance moduli have been corrected for the expansion velocity of the photosphere and the extinction derived from the plateau phase colour. Our sample of $I$ band data listed in Table \protect \ref{nir_candle} (black open circles) is compared in panel (b) to the $I$ band data of \protect \cite{poz09} (red open circles). The best-fitting line to our data is shown in each panel. The best-fitting line for \protect \cite{poz09} (red line) in the same redshift range as our data  is shown in panel (b). The formal values of the $\chi^{2}$ statistic for the $VIJHK$ bands are 10.25, 9.07, 4.28, 4.21, 1.77 respectively. However, we note that these may not be particularly meaningful given the small sample sizes. The solid black circles are SN 2002hh and SN 2004et that share the same host galaxy, NGC 6946 and the difference between their distance moduli is seen to be greatly reduced going to longer wavelengths. In panel (d), the best-fitting line excluding these two SNe is also plotted (dashed red line) and lies almost exactly on the best-fitting line for all SNe.}
\label{vel_comp}
\end{figure*}

\section{Results}
\label{results}

Using equation 2 of \cite{poz09}, we fit the absolute magnitude corrected for the 
expansion velocity and extinction using the colour excess between the \textit{V} band and the band 
being investigated. The modified equation used for the \textit{J} band for example is:

\begin{equation}
\label{eqn}
\displaystyle{M_{J} =  M_{J_0} - \alpha\ log_{10}(V_{\Feii}/5000) +R_{J}\ [(V - J) - (V-J)_{0}]         }
\end{equation}

where $R_{J}$ is based on the $R_{V}$ value of 1.5 obtained by \cite{poz09}, V$_{\Feii}$ 
is the velocity at +50 days post explosion and (\textit{V-J})$_0$ is the ridge-line colour, which we have 
taken to have a value of 1 from the average \textit{V-J} colour excess. As mentioned by \cite{nug06} 
and \cite{poz09}, this term is degenerate with M$_{J_0}$ and so its value is irrelevant. 
The fit to the data using Equation \ref{eqn} for the \textit{J} band is $\alpha$ = 6.33 $\pm$ 1.2 and  
$M_{J_0} = -18.06 \pm 0.25$ mag. While for the \textit{I} band, $\alpha$ = 4.9  $\pm$ 1.2 and 
$M_{I_0} = -17.40 \pm 0.24$ mag. These values are in agreement with those obtained by \cite{poz09} for their entire sample of 40 SNe, resulting in $\alpha$ = 4.6  $\pm$ 0.7 and  $M_{I_0} = -17.43 \pm 0.10$ mag. 
We tested the effect of varying $R_V$, and found that the commonly used value of 3.1 results in 
a worse fit across all bands with the $V$ band being most sensitive to the adopted value of $R_V$, while the scatter in the \textit{J} band increases by 0.17 mag compared to the best fit. 
For our sample of SNe, we found that $R_V$ converged to a value
of 2.0 $\pm$ 0.8. However, we opted to use $R_V = 1.5$ in our analysis as this value was derived 
from the significantly larger sample of objects \citep{poz09}. 

Figure \ref{vel_comp} shows the comparison Hubble diagram for our sample between the \textit{V}, 
\textit{I}, \textit{J}, \textit{H} and \textit{K} bands, with a scatter of 0.56, 0.50, 0.39, 0.46 and 0.40 mag in the $V$, $I$, $J$, $H$ and $K$ bands respectively.
The \textit{I} band data is first shown in combination with the data in \cite{poz09} for ease of comparison in 
Fig. \ref{vel_comp}b, while Fig. \ref{vel_comp}c shows our sample in isolation. The dispersion is dominated by the scatter in the recessional  velocities due to this sample
being relatively nearby.  Hence we find a higher dispersion in \textit{V} and \textit{I} than previously 
illustrated by \cite{ham02}, \cite{nug06} and \cite{poz09}, as one would expect due to their larger
numbers in the Hubble flow. However, in our self-consistent comparisons between the optical and 
NIR bands, we measure 
a tighter correlation in the \textit{J}  band than in \textit{I}, which appears significant. 
There does not appear to be a further improvement in going to the \textit{H} and $K$ bands. However the $K$ band, where we would expect the tightest fit, has the smallest sample size of all bands considered here, and is possibly most affected by small number statistics.
If a similar reduction in the dispersion was observed in II-P SNe in the Hubble flow, one
might expect to reduce the scatter from 0.2--0.26
\cite[as found in the \textit{I} band relations of][]{ham02,nug06,poz09},  to 
0.1--0.15.  This corresponds to distances accurate to 5--7 per cent, if the NIR photometric accuracy
can be sustained at  higher redshifts. 
The two possible physical reasons for the decreased scatter in the NIR bands are the lower effective extinction corrections 
and the relatively featureless spectra seen at NIR compared to optical wavelengths \citep{mag09}. 

\cite{poz09} had suggested that the extinction term is relatively negligible compared to the velocity correction applied in the optical, which would argue against the former reason. To investigate this further, we performed fits to the data both including and excluding
an extinction correction. Doing so, we found as expected, that the contribution to the 
extinction term is significant from $V$--$J$ bands, but that the $H$ and $K$ bands are
virtually insensitive to the application of this correction. This would suggest that the $H$ and $K$ bands should produce the tightest fits. We also tested the fits excluding the most discrepant point (SN 1999br) and this reduced the scatter across all the bands with the $K$ band, having the lowest scatter at 0.20 mag, while the $J$ and $H$ had similar values of 0.29 and 0.31 mag respectively. SN 1999br was a low luminosity SN \citep{2004MNRAS.347...74P}, 
but we find no reason to cull it from the sample and believe it would be inappropriate to do so without a sound scientific basis. These low luminosity objects are unlikely to be detectable out to high redshifts and so should not affect future samples in the Hubble flow. However, we note that larger numbers of low luminosity type II-P SNe at low redshift may skew the correlation. Future surveys will allow us to quantify the magnitude of this effect as rates of low luminosity type II-P SNe are currently not well known. Larger sample sizes are needed to clarify these results and it also leaves the open question of why with the negligible contribution from extinction in the $H$ and $K$ bands do they not produce measurably tighter fits than the $J$ band. We note that the means and standard deviations of the errors of the $JHK$ measurements are 
0.14 $\pm$ 0.13, 0.17 $\pm$ 0.13 and 0.18 $\pm$ 0.13
respectively, indicating there is no significant difference in the accuracy of the magnitudes in the NIR bands. 

A definite indicator that a much reduced extinction at NIR
wavelengths has a major role to play in the decrease in the scatter
going to longer wavelengths, is the case of SNe 2004et and 
2002hh, which share the same host galaxy. In Figure
1 these two SNe are shown as solid black circles. 
The convergence in distance moduli is improved by a factor
of $\sim$ 10 from the $V$ band to the $J$ band.

We also investigate the possibility that the scatter is reduced in the NIR bands due to the relatively few spectral features that are present in photospheric spectra. To quantify this, we have summed the equivalent widths of the spectral features above and below the continuum in spectra of SN 2004et from \cite{mag09} (i.e. the deviation from the continuum) and find that the ratio of the $J$/$I$ band features is $\sim$ 50 $\pm$ 20 per cent. The $H$/$I$ and $K$/$I$ band ratios were also determined and are $\sim$ 35 $\pm$ 20 per cent and $\sim$ 45 $\pm$ 20 per cent respectively. Thus the NIR bands have weaker spectral
features than are seen in the $I$ band region, which would imply that
photospheric temperature variations would produce less variation
in the NIR fluxes than in the optical. Further work using spectral models \citep[e.g.][]{des08} would be required to quantify.

\section{Prospects and outlook}

The immediate application of II-P SNe as distance indicators is to independently
verify cosmic acceleration between $z\simeq$ 0.3--0.5. This is of interest due to the
progenitor channels of SNe II-P and Ia being distinctly different \citep{2009ARA&A..47...63S,2000ARA&A..38..191H}, hence
one would expect different biases and effects from extinction, star formation rate and progenitor evolution. 
At $z$ of 0.3, the rest frame \textit{J} band magnitudes would be $\sim$ 23.2--25, and would
be redshifted to the $H$ band. The Wide Field Camera 3 on HST 
would produce a signal-to-noise of $\sim$ 30 with exposure times between 
 $\sim$ 1-12 h. The feeder search  would of course still be optical: the Supernova Legacy Survey and STRESS have confirmed type II SNe out to a redshift of 
$z \sim$ 0.2--0.3 \citep{2009A&A...499..653B, 2008A&A...479...49B}. 
Amongst the Pan-STARRS-1 survey's early discoveries 
is a bright type II-P at $z=0.18$ \citep[][Botticella et al., in prep.]{2009CBET.1988....1Y}
demonstrating that current surveys will harvest a sample of intermediate redshift II-P SNe. 
The velocity tracer Fe~{\sc ii} $\lambda$5169 line would be redshifted to $\sim$ 6720--7750 \AA\ and a $V$ band rest frame magnitude of 23.7--25.5, redshifted to the $RI$ bands
would mean that 8-m spectroscopy (at low-moderate dispersion) would be currently possible at $z\sim0.3$, and at 
$z\sim0.5$ for the brighter events. If the projected scatter using the standardised candle method for the rest frame 
$J$ band of 0.1--0.15 mag can be confirmed in the Hubble flow, 
then the application of distance measurements with similar accuracy to type Ia SNe is obviously
attractive. A sample of 10--15 well observed events would be sufficient to match 
the SNe Ia diagnostic ability at $z\sim$ 0.3--0.5 and confirm cosmic acceleration to 
3--4 $\sigma$ \citep{rie98,per99}. 

Future facilities such as the James Webb Space Telescope (JWST),
the European Extremely Large Telescope (E-ELT) and the proposed
US and ESA space missions JDEM and EUCLID could significantly
increase the redshift to which type II-P SNe can be observed. At $z$ of $\sim$ 0.3--0.75, the standardised candle method for type II-P SNe could be combined with measurements of baryon acoustic oscillations and the cosmic microwave background to derive joint constraints on dark energy \citep[e.g.][]{sol09}. 
Both JDEM and EUCLID baseline reference missions
target type Ia SNe in the rest frame $JH$ bands, with JDEM aiming for
1800 SNe Ia between $z\sim$ 0.3--1.2. These sensitivity limits (25.5--26 mag)
would detect II-P SNe in rest-frame $JH$ to $z\sim$ 0.75, implying
around 400 would be detected \cite[if the relative rates by volume
are similar to what we see locally;][]{sma09}. Rest-frame
optical spectroscopy  at $z$ = 0.75 is possible with the E-ELT using Laser-Tomography AO/Multi-Conjugate AO with an exposure time of
$\sim$ 1--3 h at $zJ$ band wavelengths for a signal-to-noise of 20.
Alternatively the search survey could be optical (the Large Synoptic
Survey Telescope could reach the bright events at $z\sim0.75$)
with JWST providing the NIR photometry of these events with the rest-frame
$J$ band shifted to the $K$ band with exposure times of 1--3 h. 

{{\small{ \textit{Acknowledgements}.This work, conducted as part of the award "Understanding the lives of
massive stars from birth to supernovae" (S.J. Smartt) made under the
European Heads of Research Councils and European Science Foundation
EURYI (European Young Investigator) Awards scheme, was supported by
funds from the Participating Organisations of EURYI and the EC Sixth
Framework Programme. MH ackowledges
support from FONDECYT grant 1060808, Iniciativa Cientifica Milenio
(P06-045-F), Centro de Astrofisica FONDAP (15010003), and Financiamiento
Basal CATA (PFB06). We are very grateful to V.~Ivanov for observations of some of the data presented in this paper. This research has made use of the NASA/IPAC Extragalactic Database (NED) which is operated by the Jet Propulsion Laboratory, California Institute of Technology, under contract with the National Aeronautics and Space Administration. We thank the referee, P. Nugent for comments that improved the manuscript.}}}

\end{document}